\documentclass[sigconf]{acmart}
\usepackage{threeparttable}
\usepackage{csquotes}
\usepackage{prftree}
\usepackage[normalem]{ulem}
\usepackage{xcolor}
\usepackage{cleveref}
\usepackage{enumitem}
\usepackage{units}

\newcommand{\HREF}[2]{\href{#1}{\color{blue}{#2}}}
\newcommand{\turnstile}{\vdash}
\newcommand{\del}[1]{{\color{purple}\sout{#1}}}
\newcommand{\ins}[1]{{\color{teal}#1}}
\newcommand{\Par}[1]{\noindent\textbf{#1}. }
\newcommand{\fourfifths}{\ensuremath{\nicefrac 4 5}}
\newcommand{\Rule}{\fourfifths{} rule}


\setcopyright{rightsretained}
\copyrightyear{2022}
\acmYear{2022}
\acmDOI{}
\acmISBN{}
\acmConference[]{Parity Technologies, Inc.}{Technical Report P22-1}{v0.2.2}

\begin{document}

\title{The four-fifths rule is not disparate impact}
\subtitle{A woeful tale of epistemic trespassing in algorithmic fairness}


\author{Elizabeth Anne Watkins}
\email{ew4582@princeton.edu}
\orcid{0000-0002-1434-589X}
\affiliation{%
  \institution{Princeton Center for Information Technology Policy}
  \city{Princeton}
  \state{New Jersey}
  \country{USA}
}

\author{Michael McKenna}
\email{mike@getparity.ai}
\orcid{0000-0002-8124-591X}
\affiliation{%
  \institution{Parity}
  \city{New York}
  \state{New York}
  \country{USA}
  }

\author{Jiahao Chen}
\email{jiahao@getparity.ai}
\orcid{0000-0002-4357-6574}
\affiliation{%
  \institution{Parity}
  \city{New York}
  \state{New York}
  \country{USA}
  }

\begin{abstract}
Computer scientists are trained in the art of creating abstractions that simplify and generalize. However, a premature abstraction that omits crucial contextual details creates the risk of epistemic trespassing, by falsely asserting its relevance into other contexts. We study how the field of responsible AI has created an imperfect synecdoche by abstracting the four-fifths rule (a.k.a.\ the \Rule{} or 80\% rule), a single part of disparate impact discrimination law, into the disparate impact metric. This metric incorrectly introduces a new deontic nuance and new potentials for ethical harms that were absent in the original \Rule. We also survey how the field has amplified the potential for harm in codifying the \Rule{} into popular AI fairness software toolkits. The harmful erasure of legal nuances is a wake-up call for computer scientists to self-critically re-evaluate the abstractions they create and use, particularly in the interdisciplinary field of AI ethics.
\end{abstract}



\maketitle

\begin{center}
\includegraphics[keepaspectratio,height=2in,width=\columnwidth]{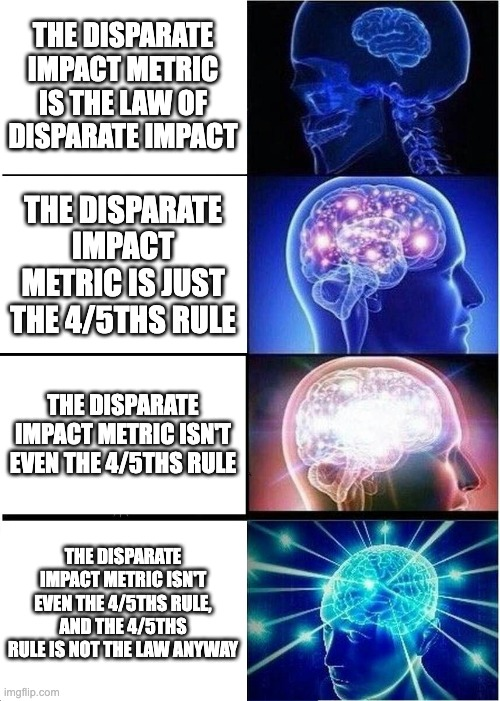}
\end{center}

\section{Introduction}

\paragraph{Premature abstraction and epistemic trespassing}
The field of computer science is oriented around two epistemic motivations:
first, to simplify complex problems into mathematical abstractions, and
second, to generalize by reusing these same abstractions across other domains \citep{Wing2006,Kramer2007}.
The creation and application of abstractions are integral to
defining computer languages and  symbolic logic in artificial intelligence
\citep{Strachey2000,sicp,castagna,Wadler2015,sep-games-abstraction}.
Abstractions discard irrelevant details,
which not only reduce cognitive load,
but also enable generalizations through use.
However, abstractions sometimes result in ontological conflicts,
particularly when the details removed in a first formulation,
especially those removed out of ignorance as to their salience
and those necessary to establish a more general context,
are regarded by others to be integral to defining the core concept in the context
from which the abstraction is constructed.
These premature abstracts, malformed through ontological errors,
cause downstream epistemic errors when reused beyond their original scope,
resulting in \enquote{research debt} \citep{research-debt}.
Well-intentioned computer scientists who lack the critical perspective
on the initial context may attempt to apply the reified abstraction
as a concept and resource in its own right,
feeling like they are simply practicing the aphorism
that \enquote{all models are wrong, but some are useful} \citep{Box1976}.
Nevertheless, such \enquote{premature abstraction} \citep{premature}--- 
using an abstraction without a critical perspective
on the original context of its creation---is problematic behavior.
By \enquote{not staying in their lane},
computer scientists can 
create semantic confusion when reborrowing the premature abstraction
back into the original context.
Rather than providing genuine contributions to the problem at hand,
they become \enquote{epistemic trespassers}, i.e.,
\enquote{thinkers who have competence or expertise to make good judgments in one field,
but move to another field where they lack competence—and pass judgment nevertheless}
\citep{ballantyne2019epistemic}.

\paragraph{Our contributions}
In this paper, we argue that epistemic trespassing has formed around the terms \enquote{disparate impact}
and \enquote{four-fifths rule},
which poses significant epistemic and deontic risks in real-world, regulated decision-making contexts.
In \Cref{sec:di-abstraction-errors},
we detail how \enquote{disparate impact} (DI\textsuperscript{law}),
a body of U.S.\ discrimination \textit{law},
and \enquote{disparate impact} (DI\textsuperscript{finding}),
a legal \textit{finding} by a court or regulator as to whether DI\textsuperscript{law} has been violated,
have been co-opted as \enquote{disparate impact} (DI\textsuperscript{metric}),
the \textit{metric} introduced into the algorithmic fairness literature
as an imperfect synecdoche of the \enquote{four-fifths rule}, which we quote in its entirety in \Cref{def:di-feldman}.
In \Cref{sec:toolkits}, we describe the spread of DI\textsuperscript{metric} in algorithmic fairness toolkits,
with societal consequences described in \Cref{sec:consequences}.
For brevity, we omit discussion of how the \Rule{} is used in regulatory compliance and enforcement \citep{eeoc89,ofccp76,ofcc79,edtest00},
focusing solely on its (lack of) validity in judicial settings.
To facilitate our discussion, we provide the relevant regulatory paragraph in \Cref{sec:reg} \citep{eeoc77,eeoc78-0,eeoc78-mar,eeoc78-cfr,eeoc79-qa,eeoc80-qa}.

\begin{definition}[Disparate impact metric (\enquote{80\% rule}, DI\textsuperscript{metric})  \citep{Feldman2015}]\label{def:di-feldman}
Given
data set $D = (X,Y, C)$, with protected attribute $X$ (e.g., race,
sex, religion, etc.), remaining attributes $Y$, and binary class to be
predicted $C$ (e.g., \enquote{will hire}), we will say that $D$ has disparate
impact if
\begin{equation}
\frac {\Pr(C=\textrm{YES} | X = 0)}
{\Pr(C=\textrm{YES} | X = 1)}
\le \tau = 0.8 \label{eq:di-metric}
\end{equation}
for positive outcome class YES and majority protected attribute
1 where $\Pr(C = c|X = x)$ denotes the conditional probability
(evaluated over $D$) that the class outcome is $c \in C$ given protected
attribute $x \in X$. 
Note that under this definition disparate impact is determined based on the given data set and decision outcomes.
\end{definition}

\paragraph{Related work}
The algorithmic fairness literature is sprinkled with various degrees of awareness of the epistemic trespassing problem around
\enquote{disparate impact}.
\citep{Feldman2015} state that \enquote{The terminology of \enquote{right}
and \enquote{wrong}, \enquote{positive} and \enquote{negative} that is used in classification is an awkward fit when dealing with majority and minority classes, and selection decisions.}
We revisit this phenomenon of deontic polarization in \Cref{sec:di-abstraction-errors}. 
Other papers focus expressly on issues around the de-/re-contextualization inherent in creating and applying abstractions.
\citep{compaslicated} comment that \enquote{Decontextualization of the data creates further problems when algorithmic fairness papers imply that their results have consequences for how [responsible AIs] work (or should work).}
\citep{selbst2019fairness} describe \enquote{the portability trap} and others
that risk creating social harms through overgeneralizations. 
\citep{martin2020extending} calls for greater community participation for creating better models and abstractions.
\citep{jacobs2021measurement} describes risks of abstracting concepts which are challenging to measure, such as gender and teacher effectiveness.
The choices of mathematical formalisms around population and data are critical to effectivel achieving fairness goals \citep{mitchell2021algorithmic}, and yet may hide harmful \enquote{methodological blindspots} with which the discipline at large must contend \citep{deng2019methodological}.
\citep{xiang2019legal} draws on theories of disparate impact in their discussion of how machine learning practitioners often misunderstand the legal concepts they attempt to operationalize.
To our knowledge, however, we are the first to provide the full synthesis of
the extent of epistemic trespassing that has happened around the terms \enquote{disparate impact}
and \enquote{four-fifths rule},
which is particularly problematic when reborrowed into the contexts of regulated decision-making
not just because of the semantic clash with DI\textsuperscript{law}, but because of the ubiquity of DI\textsuperscript{metric}.
\section{The legal concepts of disparate impact}\label{sec:di-finding}

\begin{figure*}
\centering
\includegraphics[width=\textwidth,keepaspectratio]{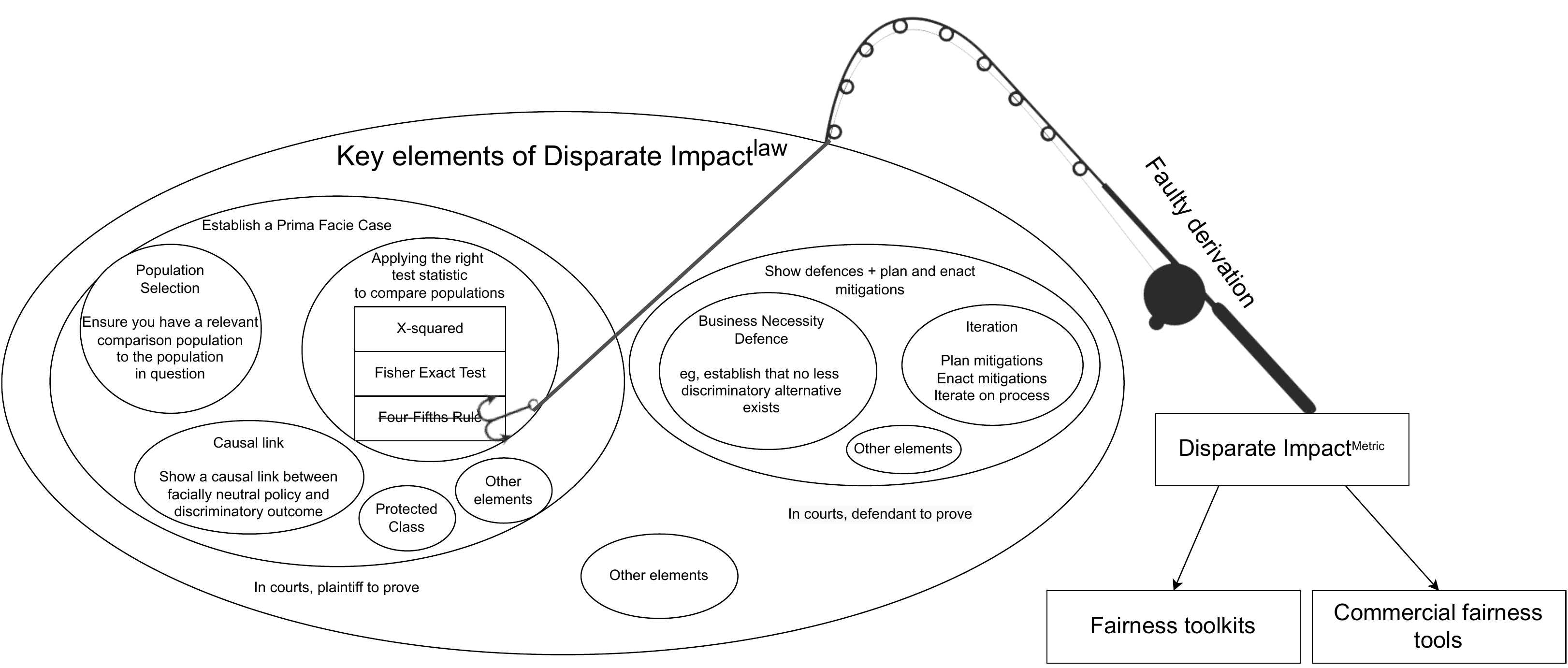}
\caption{Premature abstraction of the legal term \enquote{disparate impact} (DI\textsuperscript{law}),
showing synecdoche of the \Rule{} in the \enquote{disparate impact} metric (DI\textsuperscript{metric})
\citep{Feldman2015}.}
\label{fig:yoink}
\end{figure*}

In this section, we present the key elements needed by U.S.\ courts and regulators
to work with in the context of DI\textsuperscript{law}
to arrive at a DI\textsuperscript{finding}.
An epistemic trespasser may (falsely) presume that this is simply a matter of applying the \Rule{}
and computing DI\textsuperscript{metric} \eqref{eq:di-metric} to establish DI\textsuperscript{finding}.
They may even turn to one of the toolkits in \Cref{sec:toolkits} to perform this computation.
This chain of reasoning is an example of a fallacious synecdoche, where DI\textsuperscript{metric} vainly
stands in for the entire body of DI\textsuperscript{law}.
On the contrary, a proper DI\textsuperscript{finding} under DI\textsuperscript{law}
requires a complex iterative and multistage test, with reference to the facts of the specific case
to establish the key concepts that constitute DI\textsuperscript{law} as shown in \Cref{fig:yoink}.

\subsection{Establishing a \textit{prima facie} case of disparate impact}

The starting point for a disparate impact assessment is finding statistical evidence of a pattern of unintentional discrimination, which affects a protected class, before turning to mitigation and defences.
The resulting \textit{prima facie} case
can be established using an appropriate test statistic which compares
a relevant population 
to the specific population that is alleged to have suffered disparate impact along protected class lines,
combined with a causal link \citep{texas2015}.
For example, to assess if women suffer disparate impact in the hiring of firefighters drawing from all NYC to service Brooklyn,
a compliance team could use a $\chi^2$-test (an appropriate statistical test) to compare the $\frac{\text{women in brooklyn}}{\text{men in brooklyn}}$ or $\frac{\text{women in nyc}}{\text{men in nyc}}$ (relevant population), with $\frac{\text{women firefighters servicing brooklyn}}{\text{men firefighters servicing brooklyn}}$ (population in question)\footnote{We acknowledge the existence of genders that fall outside of the gender binary. The law typically compares against each other (one vs one), rather than comparing a class against all other classes (one vs rest)}.
If the test statistic shows a statistically significant difference, this forms evidence to be presented in court or to a regulator. 

\paragraph{Selecting a relevant comparison population}
The example above highlights an ambiguity in defining the relevant population that forms the basis for comparison when computing a test statistic.
Should the reference population be the population of New York City (as the source of applicants),
just the borough of Brooklyn (the service area), or something else?
Some cases failed to establish DI\textsuperscript{finding} because they chose too broad a reference population \citep{alexander2019}.
On the other hand, the use of general population statistics is not always inadmissible \citep{mandala2020}.
The set of relevant populations that courts will accept can turn on the legislative history as well as the facts of the case. In one recent case \citep{villafana2020},
a dispute about the appropriate comparison population drew on analogies to a range of cases in fair housing, but ultimately turned on the differences between legislative intents when writing housing and employment regulations.
Ultimately, the choice of relevant comparison population is complex, contingent, and contextual,
and cannot be easily abstracted away. 

\paragraph{Selecting an appropriate test statistic}
Once a reference population has been established, the reference population and the population under review need to be compared. In modern times, this comparison is a statistical one, but the tests that are indicated differ based on the facts of the case.
Commonly used test statistics include $\chi^2$ and Fisher's exact tests, each of which is considered reliable,
but can occasionally disagree \citep{stevenson2016}.
When conflicts between valid statistical tests arise, the court needs to make a call based on the facts of the case, as
\enquote{\enquote{[S]tatistics [...] come in infinite variety [...] their usefulness depends on all of the surrounding facts and circumstances.}}\citep{hazelwood77}
As above, the choice of testing is complex, contingent, and contextual,
and cannot be easily abstracted away. 

\paragraph{The \Rule{} is \textit{not} an appropriate test statistic}
In contrast to the tests mentioned above, the \Rule{} from which DI\textsuperscript{metric} is faultily derived is considered less favourably. \textit{It is neither necessary nor sufficient that \eqref{eq:di-metric} constitutes DI\textsuperscript{finding} in courts at all} - it is only used in employment contexts by resource-constrained regulators out of court \citep{London78,ofccp20}. Courts will simply place greater weight on significance testing than the \Rule, for reasons similar to those which inspired the exceptions in original regulation (\Cref{sec:reg}) - principally, the greater consistency of statistical significance \citep{jones2014}.

\subsection{Demonstrating a business necessity defence, or arriving at one through mitigation}

A prima facie case does not automatically lead to a DI\textsuperscript{finding}.
If significant discrepancies are found,
legal and compliance teams will look to justify the practice causing the discrepancy using business necessity justifications.
Here, the context matters.
In employment cases, it is enough to show a \enquote{nexus between its hiring requirement and the employment goals} \citep{meditz2010}. In a fair lending or machine learning context, regulators may ask for evidence that the model chosen is the least discriminatory of all models which provide sufficient value (generally, profit) \citep{elston1993}.
In a disability context, compliance teams may show that reasonable accommodations cannot rectify the alleged disparate impact\citep{payan2021}.
If mitigations are unavailable or simply too burdensome, the alleged discriminatory practice need not result in DI\textsuperscript{finding} as the above cases show. However, the discovery of mitigations and an assessment of their burden are complex matters, contingent on the facts of the case, and reliant on context.

\paragraph{Iteration}
If workable mitigations are found, they must be documented and carried out so that compliance teams can establish a business necessity defence in the future to a regulator or court. For example, if a less discriminatory alternative model is found in the process of demonstrating a business necessity defence, a bank concerned with fair lending is bound to use the less discriminatory alternative \citep{FDIC1994}. However, the less discriminatory alternative should be reassessed from the beginning, leading to an iterative process which ought to end in a process or model that can be defended in a disparate impact claim, either because no discrimination remains or because the business necessity defence can be made out.

\subsection{Summary}
The legal approaches to disparate impact analysis and mitigation are complex, expensive,
and necessary to avoid eight-digit regulatory fines, court judgments carrying similar cost, and reputational damage.
Both compliance teams and plaintiffs in court need to make subtle yet consequential choices about reference populations, statistical tests, defences, mitigation strategies, and other considerations, with reference to the particular regulatory scheme and facts of the case.
While computer scientists can help with tasks like establishing statistical evidence, there is simply no substitute for legal expertise to establish DI\textsuperscript{finding}, and DI\textsuperscript{metric} is irrelevant for DI\textsuperscript{finding}.
Computer scientists risk epistemic trespassing in overreaching for the limited places where quantitative computations are called for,
and by arguing for the synecdoche of DI\textsuperscript{metric} in place of DI\textsuperscript{finding}.

\section{Critical analysis of the generalization of disparate impact}\label{sec:di-abstraction-errors}

Having now reviewed the original legal contexts of disparate impact,
we present in this section a critical ``derivation'' of DI\textsuperscript{metric} \citep{Feldman2015}
from the regulation stating the \Rule{} (\Cref{sec:reg}).
While \Cref{def:di-feldman} claims to generalize the \Rule{} \citep{Feldman2015};
we present in \Cref{tab:abstraction_failures}
a sequence of logical transformations (introduced in \Cref{def:logic}),
showing that several premature abstractions and \textit{ad hoc} redefinitions
are necessary in this ``derivation'', which is therefore erroneous.
The flawed generalization means that
\Cref{def:di-feldman} \textit{no longer correctly describes the original regulatory use of the \Rule{}}.
To state this claim more precisely, we now introduce some formal logical definitions for the notions of premature abstraction and epistemic trespassing that we have previously introduced.
\begin{definition}\label{def:logic}
Let $\Gamma$ be some context in which the statement $x$ is true, written $\Gamma \turnstile x$;
$y$ be a statement that is more general than $x$, written $x<y$,
by virtue of omission of details;
and $\Gamma'$ be a more general context than $\Gamma$, written $\Gamma \prec \Gamma'$.
Furthermore, assume that the generality relations $<$ and $\prec$ are transitive.
Then, an \textbf{inductive generalization} (I) is the logical inference rule
\begin{displaymath}
\prftree[r]{(I)}{\Gamma \turnstile x}{\Gamma \prec \Gamma'}{x < y}{\Gamma' \turnstile y}
.
\end{displaymath}
A \textbf{decontextualization} (D) is an inductive generalization (I) where $\Gamma \prec \Gamma'$ is axiomatically presumed to be true and $x=y$ identically, i.e., only the context is asserted to be generalized and not the statement.
An \textbf{abstraction} (A) is an inductive generalization (I) where $\Gamma = \Gamma'$ identically and $x<y$ is axiomatically presumed to be true, i.e.,
only the statement is asserted to be generalized and not the context.
A \textbf{premature abstraction} (P) is an inductive generalization (I) where both $\Gamma \prec \Gamma'$ and $x<y$ are both axiomatically presumed to be true, i.e.,
both the statement and context are asserted to be generalized.
A \textbf{recontextualization} (R) is the logical inference rule
\begin{displaymath}
\prftree[r]{(R)}{\Gamma' \turnstile x}{\Gamma \prec \Gamma'}{\Gamma \turnstile x}
.
\end{displaymath}
\end{definition}

The terms de-/re-contextualization have been previously used to describe
the processes of socio-technical change \citep{Simon2006,Janneck2010}.
The reciprocal relationship turns out to be a specific instance of deontic semantics, which shows up as meaning latent in the values of binary random variables.
We now define the following concept:
\begin{definition}\label{def:dpb}
A \textbf{deontically-polarized binary (DPB) variable} is a random variable $V$ taking either a positive value or a negative value.
\end{definition}
The deontic meaning assigned to a binary variable is relevant when computing metrics of algorithmic bias.
First, \Cref{def:di-feldman} explicitly builds upon the notion of equality of outcomes, which compares base rates for the positive outcome $C=\textrm{YES}$ only (conditioned on the protected attribute).
Replacing $C=\textrm{YES}$ with $C=\textrm{NO}$ in \Cref{def:di-feldman} creates a different metric,
$\Pr(C=\textrm{NO}|X=0)/\Pr(C=\textrm{NO}|X=1)$,
which in general will not be the same value as the original ratio
$\Pr(C=\textrm{YES}|X=0)/\Pr(C=\textrm{YES}|X=1)$.
Second, the deontic value of being in the majority group $X=1$ literally defines
the denominator of the ratio: the metric is not symmetric with respect to interchanging $X=1$ and $X=0$ classes.
Third, the deontic polarizations of positive/negative outcomes ($C\in\{\textrm{YES},\textrm{NO}\}$) and majority/minority groups ($X\in\{0,1\}$) take on moral dimensions of good/bad and inclusion/exclusion that need to be considered when qualitatively assessing ethical harms.

While we do not define precisely the ``more general'' relationships $<$ or $\prec$,
the definitions above nevertheless suffice for the critique summarized in \Cref{tab:abstraction_failures}, which distinguishes between abstractions that are correct generalizations
(of the form $x<y$) and those that are not (denoted $x<^{(*)}y$),
as well as correct and incorrect logical transformations (the latter are denoted by a suffixed *).
For example, the text \enquote{A \del{selection} \ins{positive outcome} rate for any race, sex, or ethnic group which is less than four-fifths of the rate for the group with the highest rate defines disparate impact.} describes the change necessary to turn the preceding statement $x_2$ (using the word \enquote{selection}) into the current statement $x_3$ (using the phrase \enquote{positive outcome}), and similarly for mutating one context $\Gamma_i$ into another $\Gamma_{i+1}$.

\begin{table*}[ht]
\centering
\begin{tabular}{|p{0.09\textwidth}|p{0.22\textwidth}|p{0.4\linewidth}|p{0.2\textwidth}|}
\hline
Formal notation & Scope & Text & Comments \\
\hline
$\Gamma_1 \turnstile x_1$
&
Certain federal agencies and employment decisions
&
A selection rate for any race, sex, or ethnic group which is less than four-fifths of the rate for the group with the highest rate will generally be regarded by the Federal enforcement agencies as evidence of disparate impact.
&
Abridged from \Cref{sec:reg}
\\\hline
$\Gamma_2 \turnstile x_2$
&
Certain \del{federal agencies and} employment decisions
&
A selection rate for any race, sex, or ethnic group which is less than four-fifths of the rate for the group with the highest rate \del{will generally be regarded by the Federal enforcement agencies as evidence of} \ins{defines} disparate impact.
&
(P*); $\Gamma_1 \prec \Gamma_2$ discards agency; $x_1 <^{(*)} x_2$ ignores \Cref{sec:di-finding}
\\\hline
$\Gamma_3 \turnstile x_3$
&
\del{Certain employment decisions} \ins{any DPB decision
involving race, sex or ethnic groups}
&
A \del{selection} \ins{positive outcome} rate for any race, sex, or ethnic group which is less than four-fifths of the rate for the group with the highest rate defines disparate impact.
&
(P)
\\\hline
$\Gamma_4 \turnstile x_4$
&
Any DPB decision involving
\del{race, sex or ethnic groups} 
\ins{groups defined by any DPB protected attribute}
&
A positive outcome rate for any \del{race, sex, or ethnic group} \ins{binary protected attribute} which is less than four-fifths of the rate for the group with the highest rate defines disparate impact.
&
(P*); $x_3 <^{(*)} x_4$ and $\Gamma_3 \prec^{(*)} \Gamma_4$ introduce harms of categorization
\\\hline
$\Gamma_4 \turnstile x_5$
&
Any DPB decision
involving groups defined
by any DPB protected attribute
&
A positive outcome rate for any binary protected attribute which is less than four-fifths of the rate for the \del{group with the highest rate} \ins{majority group} defines disparate impact.
&
(A*); $x_4 <^{(*)} x_5$ redefines relevant population
\\\hline
$D \turnstile x_5$ &
\vspace{-6.5pt}
\ins{Data $D=(X,Y,C)$ on some} \del{any} DPB decision $C$ and \ins{some} \del{any} DPB protected attribute $X$
&
A positive outcome rate for any binary protected attribute which is less than four-fifths of the rate for the majority group defines disparate impact. &
(R) yields \Cref{def:di-feldman}
\\\hline
\end{tabular}
\caption{
Necessary abstractions to derive \Cref{def:di-feldman} from the original U.S.\ federal regulatory guidance on disparate impact. DPB is short for deontically-polarized binary variable (\Cref{def:dpb}). * denotes logically problematic steps.
See \Cref{sec:di-abstraction-errors} for details.
}
\label{tab:abstraction_failures}
\end{table*}

The logical flow of \Cref{tab:abstraction_failures}
can be summarized as a process of \textbf{epistemic trespassing},
being a fallacious premature abstraction (P*) based on faulty inductive premises
$\Gamma_1 \prec^{(*)} \Gamma_4$ and $x_1 <^{(*)} x_5$,\footnote{These statements follow from the transitive relations $\Gamma_1 \prec \Gamma_2 \prec \Gamma_3 \prec^{(*)} \Gamma_4$ and $x_1 <^{(*)} x_2 < x_3 <^{(*)} x_4 <^{(*)} x_5$.}
followed by a recontextualization (R) into the context of the data set $D$:
\begin{displaymath}
\prftree[r]{(R)}
{D\prec\Gamma_5}
{\prftree[r]{(P*)}{\Gamma_1 \turnstile x_1}{\Gamma_1 \prec^{(*)} \Gamma_4}{x_1 <^{(*)} x_5}{\Gamma_4 \turnstile x_5}}
{D \turnstile x_5}.
\end{displaymath}
While the second step is logically valid,
the first step involves problematic assertions which invalidate the premises
upon which the premature abstraction was defined.

The individual steps reveal the precise logical faults worth detailing,
as are the concomitant implicit, yet necessary, widenings of context
to enable abstracting away of now-irrelevant details.
$\Gamma_1 \prec \Gamma_2$ enables the notion of (enforcement) agency,
to be discarded in $x_1 <^{(*)} x_2$, which now claims the \Rule{} as an operational definition,
and glosses over all the legal requirements described in \Cref{sec:di-finding}.
The generalization $x_2 < x_3$, while abstracting away the binary decision,
needs to retain the deontic polarity presumed in that it is a good thing for people to be employed,
which must be preserved in the widening $\Gamma_2 \prec \Gamma_3$ even without the employment context.
The authors of \citep{Feldman2015} have acknowledged such deontic polarity as \enquote{awkward}.
The generalization $x_3 <^{(*)} x_4$
abstracts away \enquote{protected attributes},
a reference to the legal notion of protected class in DI\textsuperscript{law}.
However, the \enquote{binary} modifier collapses nuance in the comparisons to be measured
and introduces an ethical harm of overly broad categorization,
making the widening of the corresponding context $\Gamma_3 \prec^{(*)} \Gamma_4$ problematic.
For example, rather than considering each racial group separately relative to some reference racial group,
the nuance is flattened into a simple pairwise comparison of out-group performance relative to in-group performance,
and lays bare the deontic subtext that belies the comparison.
This problem is further exacerbated in the transformation
$x_4 <^{(*)} x_5$, which redefines the reference group
as the majority group $X=1$.
The only way to view this redefinition as an abstraction
is to assume that \textit{in all reference populations, the majority group and most advantaged group are identical}.
This change in semantics alters the description of model minorities that are not the majority group ($X=0$) but are nevertheless the group more likely to have the better outcomes, $\Pr(C=\textrm{YES}|X=0) > \Pr(C=\textrm{YES}|X=1)$,
and \textit{decouples the deontic polarity of the outcome $C$ from the deontic polarity of the in-group membership $X$},
requiring now the management of two separate sets of deontic semantics.
All these semantic changes culminate in the recontextualization going from $\Gamma_4$ to an empirical data set $D$,
which requires categorical assignments into the DPB variables $X$ and $C$ in order to be defined.
Furthermore, the rows of $D$ explicitly define the reference population to be assessed,
raising practical issues around representativeness and sampling bias that must be considered.

\subsection{Possibilities for removing deontic polarization}

The analysis above demonstrate the composition of multiple abstractions that were necessary to arrive at \Cref{def:di-feldman}.
It is also clear that other abstractions of the \Rule{} are also possible,
being analogous to \eqref{def:di-feldman} as a codification of $D \turnstile x_5$,
albeit corresponding to statements other than $x_5$.
Thus, it is possible to ameliorate one of the most problematic aspects of DI\textsuperscript{metric} by redefining the test to
remove the deontic aspect of the protected attribute $X$.
For example, $\Gamma_4 \turnstile x_4$ could have been 
codified for a specific data set, $D \turnstile x_4$, as the symmetrized ratio
\begin{equation}
\min\left( \frac {\Pr(C=\textrm{YES} | X = 0)}
{\Pr(C=\textrm{YES} | X = 1)}, \frac {\Pr(C=\textrm{YES} | X = 1)}
{\Pr(C=\textrm{YES} | X = 0)} \right)
\le \tau = 0.8,   
\end{equation}
or equivalently,
\begin{equation}
\tau \le \frac {\Pr(C=\textrm{YES} | X = 0)}
{\Pr(C=\textrm{YES} | X = 1)} \le \frac 1 {1-\tau}.
\label{eq:di-metric-sym}
\end{equation}
This redefinition removes deontic polarization by
symmetrization: it no longer matters which group $X=1$ or $X=0$
serves as the basis for comparison and goes into the denominator.
As we will see later in \Cref{sec:toolkits}, some practical implementations
of DI\textsuperscript{metric} do indeed 
define the \Rule{} in terms of \eqref{eq:di-metric-sym} instead of \eqref{eq:di-metric}.

An alternative to removing the deontic polarization of $X$ is to
attempt a different way to abstract away the
\enquote{race, sex or ethnic group} of $\Gamma_3$ and $x_3$,
which assigns to $X$ the deontic meaning of $C$ rather than
giving $X$ its own, separate, deontic semantics.
For example, consider
$\Gamma_{4'} \turnstile x_{4'}$:

\noindent\begin{tabular}{|p{0.3\columnwidth}|p{0.625\columnwidth}|}
\hline
Any DPB decision involving a \ins{categorical protected attribute}
&
A positive outcome rate for any \ins{categorical protected attribute} which is less than four-fifths of the rate for the group with the highest rate defines disparate impact.
\\\hline
\end{tabular}
which could have been codified $D\turnstile x_{4'}$ into the ratios
\begin{equation}
\rho^{(4')}(x) = \frac {\Pr(C=\textrm{YES} | X = x)}
{\max_{x'} \Pr(C=\textrm{YES} | X = x')}
\le \tau, \textrm{where } x\ne x',
\end{equation}
where the denominator encodes the notion of \enquote{group with the highest rate}
and not \enquote{majority group}.
A single metric could have been constructed from summary statistics of these ratios;
one plausible metric, $R^{(4')}$, is simply to consider the worst case:
\begin{equation}
R^{(4')} = \min_x \rho^{(4')}(x). \label{eq:di-4'}
\end{equation}
In the special case of a binary $X$ (with no deontic polarization needed),
\eqref{eq:di-4'} reduces to \eqref{eq:di-metric-sym}; it is therefore accurate to characterize
\eqref{eq:di-4'} as the correct generalization of \eqref{eq:di-metric-sym} to categorical $X$.

While the above it is possible to remove the deontic polarization necessary in $X$, the preceding discussion also shows how it is impossible to completely remove deontic polarization from any redefinition of DI\textsuperscript{metric}, for two reasons.
First, the deontic polarization of $C$ is necessary for correctly computing DI\textsuperscript{metric}.
Consider the confusion matrix
\begin{tabular}{|c|cc|}
\hline
      & $C=1$ & $C=0$ \\\hline
$X=1$ & $P_1$ & $N_1$ \\
$X=0$ & $P_0$ & $N_0$ \\
\hline
\end{tabular} ,
where $P_x$ is the number of people receiving the positive outcome $C=1$ that belong to $X=x$,
and $N_x$ being the corresponding negative count. A simple computation of DI\textsuperscript{metric}
yields the ratio 
$(1 + \nicefrac{N_1}{P_1})/(1 + \nicefrac{N_0}{P_0})$.
If $C=0$ were the positive outcome, DI\textsuperscript{metric} would instead take the reciprocal value
$(1 + \nicefrac{N_0}{P_0})/(1 + \nicefrac{N_1}{P_1})$.
Second, the assumption of universal positive polarity in $C=1$ neglects more complex nuances;
in the original context of employment, the holistic consideration of
the underemployment of women \citep{Weststar2011}, youths \citep{Churchill2021}, and racial minorities \citep{Nelson2021};
exploitative labor conditions that affect vulnerable workers like those in lower-income countries \citep{Snyder2010}, children \citep{Radfar2018}, and trafficked slaves \citep{Weitzer2015,Patterson2018};
and other concerns around people with disabilities \citep{Miethlich2019,Sauer2010},
social class \citep{Roemer1982},
immigration \citep{Okkerse2008},
labor organizing \citep{Sachs2007},
freelancing \citep{Sutherland2020},
and corporate social responsibility \citep{Etter2019},
are all necessary for determining the deontic value of an employment selection.
Similar deontic assumptions must be confronted in other contexts, such as granting bail to those who cannot afford it \citep{compaslicated}.

These possibilities for ameliorating a single problematic aspect of DI\textsuperscript{metric},
while instructive for understanding how to improve the quantitative definition,
nevertheless do not redress all the various stages of premature abstraction that
enable the epistemic trespassing of the \Rule.
Rather, this discussion lays bare how far the meaning of DI\textsuperscript{metric} has deviated from the original meaning of the \Rule{} in DI\textsuperscript{law}.
Therefore, we argue that giving the name \enquote{disparate impact} to DI\textsuperscript{metric} is a problematic practice,
and exhort the algorithmic fairness community to stop this usage of \enquote{disparate impact}, to reduce the inevitable epistemic trespassing in conflating DI\textsuperscript{metric} with DI\textsuperscript{finding}.
Worryingly, such epistemic trespassing is already manifest in AI fairness toolkits,
as we will now discuss.

\section{Spreading the \Rule{} in fairness toolkits}\label{sec:toolkits}

Interest in fairness and disparate impact within the computer science discipline has grown greatly since \citep{Feldman2015} was published in 2015.
Perhaps in response to this growing demand for applicable fairness heuristics which can be implemented into statistical models, a new field has emerged of \enquote{AI ethics} and \enquote{AI fairness} toolkits. Such toolkits are usually open-source code, but commercial offerings do exist. These technical packages operationalize guidelines for \enquote{fair} decision-making into tests which end-users can build into their own model-development processes to assess their own models' treatment of disparate groups, or use as-is. 

A number of papers have critiqued the presumptions and organizational imperatives of toolkits, in particular how these toolkits prioritize the decision-making of privileged technologists \citep{neff2020bad}, how they frame the work of AI ethics as an individual rather than systematic endeavor \citep{merrill_madaio_wong_2022} and may fail to address practitioner needs \citep{richardson2021towards,lee2021landscape}.
Rather, we focus on toolkits as constructions that collect instruments, processes, and actions in prescriptive ways
that make a deliberate representation of expertise \citep{mattern_2021}.
Fairness toolkits perpetuate the epistemic trespassing we have detailed above in \Cref{sec:di-abstraction-errors}, which lends undue weight to DI\textsuperscript{metric} by giving it the same name as DI\textsuperscript{law}.
These toolkits are clearly not built to handle the full complexities of DI\textsuperscript{law} as sketched in \Cref{fig:yoink},
and since few, if any, users of these toolkits will be aware of the nuances of disparate impact that we differentiate in this paper,
offering DI\textsuperscript{metric} under a name like \enquote{disparate impact} ought to provoke concern about unintended legal claims that are unwarranted from simply computing DI\textsuperscript{metric}. Overall, the inadequacy of toolkits to assure legal protections, combined with their widespread popularity, makes our argument both compelling and urgent. 


\subsection{Fairness toolkits are popular}

\begin{table*}[ht]
\begin{threeparttable}
\begin{tabular}{|p{0.25\textwidth}|p{0.1\textwidth}|p{0.1\textwidth}|p{0.1\textwidth}|p{0.2\textwidth}|p{0.1\textwidth}|}
\hline
Name      & GitHub stars & PyPI downloads & Paper citations\tnote{1}
 & Name of DI metric or similar & Suggests an 80\% threshold \\\hline
Aequitas  & 458                              & 83750                              & 106                                 & Impact parity\tnote{2}               & Yes\tnote{3} \\
AIF360    & 1635                             & 178736                             & 340                                 & \HREF
{https://aif360.readthedocs.io/en/latest/modules/generated/aif360.sklearn.metrics.disparate_impact_ratio.html}
{{Disparate impact ratio}}      & Yes                        \\
Fairlearn & 1190                             & 391898                             & 58                                  & Selection rate ratio         & No                         \\
Audit-AI  & 273                              & 21159                              & N/A        & 4/5 test                     & Yes                       \\
Salesforce Einstein  &          N/A                  &       N/A                        & N/A        & Disparate impact                     & Yes                       \\
Fairplay Mortgage Fairness Monitor  &          N/A                     &                N/A               & N/A        & Adverse impact ratio                   & Yes                       \\
H2O.ai &          N/A                     &                N/A               & N/A        & Adverse impact ratio                   & Yes                       \\
\hline
\end{tabular}
\begin{tablenotes}
\item[1] Citation counts taken from Google Scholar.
\item[2] Documentation also refers to 
DI\textsuperscript{metric} as \enquote{proportional parity} or \enquote{minimizing disparate impact}.
\item[3] Documentation recommends the 80\% threshold not just for DI\textsuperscript{metric}, but for multiple similar metrics.
\end{tablenotes}
\end{threeparttable}
\caption{Statistics of popularity for several major AI fairness toolkits as of 2022-02-16.
}
\label{tab:toolkits}
\end{table*}

In this section, we briefly overview some AI fairness toolkits that present functionality for computing bias metrics,
and highlight any references made to \enquote{disparate impact} or the \Rule{} in their documentation.
\Cref{tab:toolkits} shows some crude statistics that indicate the relative popularity of each toolkit.
Altogether, these toolkits have been downloaded at least 600,000 times,
which ought to raise concerns about the scale of unintended and improper legal claims
that may be perpetrated across all sorts of use cases.

\begin{description}[leftmargin=0pt, align=left] 

\item[Microsoft Fairlearn \HREF{https://pepy.tech/project/fairlearn}{(390,000+ downloads)}] \citep{fairlearn}
is the only
fairness toolkit we surveyed here which does not use \enquote{Disparate Impact} in its naming of DI\textsuperscript{metric}, and also does not suggest any thresholds (in particular those that align with the \Rule).
Furthermore, Fairlearn's documentation acknowledges risks inherent in \enquote{portability traps} and the like.
We commend the authors of Fairlearn for their care in avoiding epistemic trespassing.\footnote{We found an example where a data scientist could not find suggested thresholds in Fairlearn, and so looked to the thresholds in AI Fairness 360, found 80\% thresholds and ended up using the \Rule{} anyway \citep{prog.world}.}

\item[Aequitas \HREF{https://pepy.tech/project/aequitas}{(80,000+ downloads)}] \citep{aequitas}
relies heavily on the 0.8--1.25 thresholds which characterize \eqref{eq:di-metric-sym}, and in fact exhibits additional epistemic trespassing by applying these thresholds to metrics other than DI\textsuperscript{metric}.
For instance, the \HREF{https://dssg.github.io/aequitas/output_data.html}{main example for their Bias Report} states that
\enquote{any disparity measure between 0.8 and 1.25 will be deemed fair. (This is inline with the 80 percent rule for determining disparate impact).}
The \HREF{http://aequitas.dssg.io/example.html}{corresponding report} claims that meeting the \Rule{} will ensure a \enquote{pass} grade for the audit: \enquote{If disparity for a group is within 80 percent and 125 percent of the value of the reference group on a group metric (e.g. False Positive Rate), this audit will pass.}

\item[pymetrics Audit-AI \HREF{https://pepy.tech/project/audit-ai}{(20,000+ downloads)}]
has a \HREF{https://github.com/pymetrics/audit-ai}{README} explicitly cites EEOC and the \Rule.
They then provide a sample model problem describing a ratio of a \enquote{lowest-passing} population to the \enquote{highest-passing} population, describing a \enquote{ratio [that] is greater than .80 (4/5ths), the legal requirement enforced by the EEOC, the model would pass the check for practical significance.} While the author takes care to denote that the EEOC guidelines originate in the hiring space, they explicitly generalize the rule to all domains (including outside employment) without warning users that different rules may apply.

\item[IBM AI Fairness 360 (AIF360, \HREF{https://pepy.tech/project/aif-360}{175,000+ downloads})] \citep{aif360}
depicts the \fourfifths{} threshold in their \HREF{https://aif360.mybluemix.net}{GUI tutorial}.
In their \HREF{https://github.com/Trusted-AI/AIF360/blob/master/examples/tutorial_medical_expenditure.ipynb}{notebook tutorial} on a medical expenditure data set, they note that \enquote{$1-\min(DI, 1/DI) < 0.2$ is typically desired for classifier predictions to be fair}, which is equivalent to \eqref{eq:di-metric-sym}.

\item[Salesforce's Einstein] is a proprietary tool including bias safeguarding, which \HREF{https://res.cloudinary.com/hy4kyit2a/f_auto,fl_lossy,q_70/learn/modules/ethical-model-development-in-einstein-discovery-quick-look/use-einstein-discovery-to-detect-and-prevent-bias-in-models/images/7375c53432d53796a5d61226f41c1a40_a-6487-f-98-2-fd-7-421-d-bf-0-b-48-f-5527-e-710-c.png}{depicts} the four-fifths threshold for DI\textsuperscript{metric} in their demo under the name \enquote{Disparate Impact}. Their customer story indicates that Einstein is used in a finance context, where particularly onerous anti-discrimination law applies.

\item[Fairplay AI's Mortgage Fairness Monitor] is a proprietary tool which measures mortgage fairness by county. The \HREF{https://fairplay.ai/mortgage-fairness-monitor/}{tool} uses DI\textsuperscript{metric}, termed Adverse Impact Ratio. The thresholds used are <80\%, between 80\% and 90\%, and over 90\%. Their target market is finance, where particularly onerous anti-discrimination law applies.

\item[H2O.ai] offers a responsible ML workflow paper \citep{h2o-responsibleml}, which acknowledges that \enquote{it is not clear that the use of this [80\%]
threshold is directly relevant to testing fairness for measures other than the AIR.} A \HREF{https://www.h2o.ai/blog/mitigating-bias-in-ai-ml-models-with-disparate-impact-analysis/}{\textcolor{blue}{blog post}} which describes \enquote{Disparate Impact Analysis}
or DIA, states that \enquote{The regulatory agencies will generally regard a selection rate for any group which is less than four-fifths (4/5) or eighty percent of the rate for the group with the highest selection rate as constituting evidence of adverse impact} immediately following the sentence of describing \enquote{discrimination in hiring, housing, etc., or in general any public policy decisions}, which can be read as epistemic trespassing in claiming the relevance of the \Rule{} in all public policy decisions.
The same post claims that \enquote{Disparate Impact Analysis is one of the tools that is broadly applicable to a wide variety of use cases under the regulatory compliance umbrella, especially around intentional discrimination.} On the contrary, intent is irrelevant to establishing DI\textsuperscript{finding}. Other \HREF{https://github.com/h2oai/driverlessai-tutorials/blob/master/compliant_driverlessai/notebooks/compliant_dia_gender.ipynb}{\textcolor{blue}{tutorials}} also explicitly reference the same 0.8--1.25 range of \eqref{eq:di-metric-sym} to \enquote{be flagged as disparate.}
\end{description}

\section{Risks from epistemic trespassing of the \Rule}\label{sec:consequences}

The use of the DI\textsuperscript{metric} as a stand-in for DI\textsuperscript{law} carries obvious legal risks for users.
Not only does this synecdoche gloss over all the requirements of \Cref{sec:di-finding},
the decontextualization $x_1 < x_3$ falsely presents DI\textsuperscript{metric} as relevant outside U.S.\ federal employment law,
when in reality, no evidence exists for its adoption into other domains.
In contrast, many toolkits in \Cref{sec:toolkits} encourage this epistemic trespassing, creating a self-fulfilling prophecy
of relevance spillover, not just into other U.S.\ regulatory contexts, but even into non-U.S.\ jurisdictions \citep{sanchez2019does}!
Furthermore, the conflation of DI\textsuperscript{metrics} with DI\textsuperscript{finding} creates a slippery slope of presuming that if \eqref{eq:di-metric} constitutes unfairness, then debiasing is simply a matter of transforming the data $D$ so that \eqref{eq:di-metric} is now falsified. Such a misconception underlies the presentation of the disparate impact remover technique \citep{Feldman2015} that is again available in many of the toolkits of \Cref{sec:toolkits}. A proper approach to debiasing is an important matter for future research.

\section{Conclusion and outlook}

While everyone wants computer systems to not discriminate,
reaching for a single DI\textsuperscript{metric} to encompass the entire body of DI\textsuperscript{law}
is overly reductive and trivializes important aspects of establishing DI\textsuperscript{finding}.
The epistemic trespassing inherent in conflating all of these DIs does a disservice
to real-world decision-making systems that must operate in regulatory contexts where DI\textsuperscript{law} applies,
and is unfortunately manifest in multiple, popular software toolkits.

The very real potential for causing harm through well-intentioned misuse of these toolkits
requires computer scientists to be more self-critical in their zeal for abstraction,
and to be willing to revise initial abstractions when ontological errors in their formation are later elucidated.
The self-awareness of the limitations of computational thinking via abstractions
is essential for working across disciplinary boundaries, particularly with lawyers,
who primarily reason by analogy to specific cases and appeals to authority.
Such self-criticism will be essential for incrementally improving upon the practice of \textit{ethical} decision-making,
around which awareness on processes like checklists, model cards, and datasheets is
emerging \citep{madaio2020co,gebru2021datasheets,mitchell2019model}. 

\begin{acks}
We thank Liz O'Sullivan and Aleksander Eskilson for helpful discussions,
and Rumman Chowdhury for feedback on an earlier draft.
The meme on the front page was created on \HREF{https://imgflip.com}{imgflip.com}
\end{acks}

\bibliographystyle{ACM-Reference-Format}
\bibliography{bib}

\appendix

\section{The \Rule{} in regulation}\label{sec:reg}
For ease of reference, we quote verbatim the entire paragraph from the U.S.\ Code of Federal Regulations that describes the \Rule.
This paragraph forms part of the Uniform Guidelines on Employee Selection Procedures (29 CFR \S1607).
We ignore the minor legal nuance that distinguishes disparate impact (DI\textsuperscript{finding}) from adverse impact, and treat them synonymously.

\begin{displayquote}
\Par{29 CFR \S1607.4(D) Adverse impact and the \enquote{four-fifths rule}}.
A selection rate for any race, sex, or ethnic group which is less than four-fifths (4/5) (or eighty percent) of the rate for the group with the highest rate will generally be regarded by the Federal enforcement agencies as evidence of adverse impact, while a greater than four-fifths rate will generally not be regarded by Federal enforcement agencies as evidence of adverse impact.
Smaller differences in selection rate may nevertheless constitute adverse impact, where they are significant in both statistical and practical terms or where a user's actions have discouraged applicants disproportionately on grounds of race, sex, or ethnic group. Greater differences in selection rate may not constitute adverse impact where the differences are based on small numbers and are not statistically significant, or where special recruiting or other programs cause the pool of minority or female candidates to be atypical of the normal pool of applicants from that group. Where the user's evidence concerning the impact of a selection procedure indicates adverse impact but is based upon numbers which are too small to be reliable, evidence concerning the impact of the procedure over a longer period of time and/or evidence concerning the impact which the selection procedure had when used in the same manner in similar circumstances elsewhere may be considered in determining adverse impact. Where the user has not maintained data on adverse impact as required by the documentation section of applicable guidelines, the Federal enforcement agencies may draw an inference of adverse impact of the selection process from the failure of the user to maintain such data, if the user has an underutilization of a group in the job category, as compared to the group's representation in the relevant labor market or, in the case of jobs filled from within, the applicable work force.
\end{displayquote}

The legal scope of this paragraph is defined in an earlier section, which we also quote verbatim for ease of reference and to illustrate the full complexity of the legal scope in which the \Rule{} is defined.

\begin{displayquote}
\Par{29 CFR \S1607.2 Scope}

\Par{A. Application of guidelines}
These guidelines will be applied by the Equal Employment Opportunity Commission in the enforcement of title VII of the Civil Rights Act of 1964, as amended by the Equal Employment Opportunity Act of 1972 (hereinafter “title VII”); by the Department of Labor, and the contract compliance agencies until the transfer of authority contemplated by the President's Reorganization Plan No. 1 of 1978, in the administration and enforcement of Executive Order 11246, as amended by Executive Order 11375 (hereinafter “Executive Order 11246”); by the Civil Service Commission and other Federal agencies subject to section 717 of title VII; by the Civil Service Commission in exercising its responsibilities toward State and local governments under section 208(b)(1) of the Intergovernmental-Personnel Act; by the Department of Justice in exercising its responsibilities under Federal law; by the Office of Revenue Sharing of the Department of the Treasury under the State and Local Fiscal Assistance Act of 1972, as amended; and by any other Federal agency which adopts them.

\Par{B. Employment decisions}
These guidelines apply to tests and other selection procedures which are used as a basis for any employment decision. Employment decisions include but are not limited to hiring, promotion, demotion, membership (for example, in a labor organization), referral, retention, and licensing and certification, to the extent that licensing and certification may be covered by Federal equal employment opportunity law. Other selection decisions, such as selection for training or transfer, may also be considered employment decisions if they lead to any of the decisions listed above.

\Par{C. Selection procedures}
These guidelines apply only to selection procedures which are used as a basis for making employment decisions. For example, the use of recruiting procedures designed to attract members of a particular race, sex, or ethnic group, which were previously denied employment opportunities or which are currently underutilized, may be necessary to bring an employer into compliance with Federal law, and is frequently an essential element of any effective affirmative action program; but recruitment practices are not considered by these guidelines to be selection procedures. Similarly, these guidelines do not pertain to the question of the lawfulness of a seniority system within the meaning of section 703(h), Executive Order 11246 or other provisions of Federal law or regulation, except to the extent that such systems utilize selection procedures to determine qualifications or abilities to perform the job. Nothing in these guidelines is intended or should be interpreted as discouraging the use of a selection procedure for the purpose of determining qualifications or for the purpose of selection on the basis of relative qualifications, if the selection procedure had been validated in accord with these guidelines for each such purpose for which it is to be used.

\Par{D. Limitations}
These guidelines apply only to persons subject to title VII, Executive Order 11246, or other equal employment opportunity requirements of Federal law. These guidelines do not apply to responsibilities under the Age Discrimination in Employment Act of 1967, as amended, not to discriminate on the basis of age, or under sections 501, 503, and 504 of the Rehabilitation Act of 1973, not to discriminate on the basis of disability.

\Par{E. Indian preference not affected}
These guidelines do not restrict any obligation imposed or right granted by Federal law to users to extend a preference in employment to Indians living on or near an Indian reservation in connection with employment opportunities on or near an Indian reservation.
\end{displayquote}

\end{document}